\documentclass[11pt,citesort]{article}

\usepackage{geometry}
\usepackage{axodraw}
\newcommand{\ice}[1]{\relax}
\usepackage{graphicx}
\usepackage{epsfig}
\usepackage{latexsym}
\usepackage{amsmath}
\usepackage{amsfonts}
\usepackage{amsxtra}
\usepackage{cite}

\def\krto{ {\,\,\lower .8ex\hbox {$\longrightarrow \atop k \rightarrow 0$}\,\,}}
\def\Section#1{\section{#1}\hspace{\parindent}}

\def\bea{\begin{eqnarray} }
\def\beq{\begin{eqnarray} }

\def\eea{\end{eqnarray}}
\def\eeq{\end{eqnarray}}
\def\eq#1{eq.~(\ref{#1})}

\newcommand{\ghostSD}{\begin{picture}(150,25)(0,0)
\SetWidth{1.2}
\DashArrowLine(12.5,0)(37.5,0){5}
\DashArrowLine(37.5,0)(75,0){5}
\DashLine(75,0)(112.5,0){5}
\DashArrowLine(112.5,0)(137.5,0){5}
\SetWidth{1}
\Vertex(112.5,0){2}
\GlueArc(75,0)(37.5,0,90){-4}{6}
\GlueArc(75,0)(37.5,90,180){-4}{6}
\CCirc(75,0){5}{Black}{Yellow}
\CCirc(75,37.5){5}{Black}{Yellow}
\CCirc(37.5,0){5}{Black}{Yellow}
\Text(20,-10)[l]{a,k}
\Text(50,15)[l]{d,$\nu$}
\Text(100,-10)[l]{e}
\Text(100,15)[r]{f,$\mu$}
\Text(50,-10)[l]{c,q}
\Text(120,-10)[l]{b,k}
\Text(75,48)[c]{q-k}
\end{picture}}

\newcommand{\ghostDr}{\begin{picture}(100,25)(0,0)
\SetWidth{1.2}
\DashArrowLine(12.5,0)(50,0){5}
\DashArrowLine(50,0)(87.5,0){5}
\CCirc(50,0){5}{Black}{Yellow}
\Text(12.5,-10)[l]{a}
\Text(87.5,-10)[r]{b}
\Text(50,-10)[c]{k}
\end{picture}}

\newcommand{\ghostBr}{\begin{picture}(100,25)(0,0)
\SetWidth{1.2}
\DashArrowLine(12.5,0)(87.5,0){5}
\Text(12.5,-10)[l]{a}
\Text(87.5,-10)[r]{b}
\Text(50,-10)[c]{k}
\end{picture}}

\begin{document} 
\date{}

\title{ The low-momentum ghost dressing function and the gluon mass }
\author{ Ph. Boucaud$^a$, M.~E. G\'omez$^b$, J.P. Leroy$^a$, A. Le Yaouanc$^a$, J.~Micheli$^a$, 
\\ O. P\`ene$^a$, J. Rodr\'{\i}guez-Quintero$^b$}

\maketitle

\begin{center}
$^a$Laboratoire de Physique Th\'eorique\footnote{Unit\'e Mixte de Recherche 8627 du Centre National de 
la Recherche Scientifique} \\
Universit\'e de Paris XI, B\^atiment 211, 91405 Orsay Cedex,
France \\
$^b$ Dpto. F\'isica Aplicada, Fac. Ciencias Experimentales,\\
Universidad de Huelva, 21071 Huelva, Spain.
\end{center}

\begin{abstract}

We study the low-momentum ghost propagator Dyson-Schwinger equation (DSE) in Landau gauge,
assuming for the truncation a constant ghost-gluon vertex, as it is extensively done, and 
a simple model for a massive gluon propagator. Then, regular DSE solutions (the zero-momentum 
ghost dressing function not diverging) appear to emerge and we show the ghost propagator to be 
described by an asymptotic expression reliable up to the order ${\cal O}(q^2)$. 
That expression, depending on the gluon mass and the zero-momentum Taylor-scheme 
effective charge, is proven to fit pretty well 
the low-momentum ghost propagator obtained through big-volume lattice 
simulations.

\end{abstract}

\begin{flushright}
{\small UHU-FP/10-023}\\
{\small LPT-Orsay/10-29}\\
\end{flushright}




\Section{Introduction}

A few years ago (see for instance \cite{Alkofer:2000wg}), a vanishing gluon propagator and a diverging ghost 
dressing function at zero-momentum in Landau gauge 
were extensively accepted as the solutions for the tower of Dyson-Schwinger equations (DSE). 
In contrast, alternative DSE solutions were also predicted to give a massive gluon 
propagator~\cite{Aguilar:2006gr,Aguilar:2008xm}. Lattice QCD (LQCD) estimates for those propagators appeared 
to be also in contradiction with a gluon propagator that vanishes at zero-momentum or with a 
ghost dressing function that diverges~\cite{Cucchieri:2007md,Bogolubsky:2007ud,IlgenGrib,Boucaud:2005ce}.
We addressed this issue in two recent papers~\cite{Boucaud:2008ji,Boucaud:2008ky} and tried to clarify 
the contradiction. After assuming in the vanishing momentum limit a ghost dressing function behaving as 
$F(q^2) \sim (q^2)^{\alpha_F}$ and a gluon propagator as 
$\Delta(q^2) \sim (q^2)^{\alpha_G-1}$ (or, by following a notation commonly used,
a gluon dressing function as $G(q^2)= q^2 \Delta(q^2) \sim (q^2)^{\alpha_G}$), 
we proved that the ghost propagator DSE (GPDSE) admits two types of solutions:

\begin{itemize}
\item If $\alpha_F \neq 0$, the low-momentum behaviour of both gluon and ghost propagators 
are related by the condition $2 \alpha_F+\alpha_G = 0$ implying that $F^2(q^2)G(q^2)$ goes to a
non-vanishing constant when $q^2 \to 0$. 
\item If $\alpha_F=0$, the low-momentum leading term of the gluon propagator is constrained not any longer 
by the leading but instead by the next-to-leading one of the ghost propagator, and LQCD 
solutions indicating that $F^2(q^2)G(q^2) \to 0$  when $q^2 \to0$~\cite{IlgenGrib,Boucaud:2005ce} 
can be pretty well accomodated within this case. 
\end{itemize}

In particular, the numerical study in ref.~\cite{Boucaud:2008ji} of the GPDSE using a LQCD 
gluon input finds that both cases of solutions appear depending on the value of the
strong coupling constant at the renormalization point, which is a free parameter 
in this exercise. Indeed, it seems to be by now well established that the two classes 
of solutions, dubbed {\it ``decoupling''} ($\alpha_F=0$) and {\it ``scaling''} 
($\alpha_F \neq 0$) may emerge from the tower of DSE~\cite{Fischer:2008uz,Aguilar:2006gr,Aguilar:2008xm}.
Such a nomenclature, despite being widely accepted, 
can be misleading. The perturbative running for the 
coupling constant renormalized in Taylor-scheme is given by $F^2(q^2)G(q^2)$ and one 
can thus define, although not univocally in the IR, a coupling with it. 
Nevertheless, neither a scale invariance nor a decoupling of the IR dynamics for the theory 
can be inferred  from the low-momentum behaviour of such a coupling. In particular, as will be seen, 
an effective charge can be properly defined for phenomenological purposes such that it reaches a 
constant at zero-momentum in the decoupling case. However, although not appropriate for 
phenomenological purposes in the IR domain, the Taylor-scheme coupling is a very convenient quantity in 
discriminating the kind of solutions we deal with.
 
On the other hand, it was also proved in ref.~\cite{Boucaud:2008ji} that, for an 
appropriate coupling constant value at the renormalization momentum, the
resulting ghost dressing function (belonging to the decoupling class) fits very well with
lattice results. It is worth pointing too that 
lattice data can be also very well accomadated within DS coupled 
equations in the PT-BFM scheme~\cite{Aguilar:2006gr,Aguilar:2008xm} and within 
the Gribov-Zwanziger\footnote{In addition, K-I. Kondo triggered very recently an interesting 
discussion about the Gribov horizon condition and its implications on the Landau-gauge 
Yang-Mills infrared solutions~\cite{Kondo:2009ug,Boucaud:2009sd}.} 
approach~\cite{Dudal:2007cw}, leading in both cases to 
decoupling solutions for gluon and ghost propagators.

Furthermore, in ref.~\cite{Boucaud:2008ky}, the low-momentum first correction to the constant leading behaviour 
of the ghost dressing function for the decoupling solution was proven to be proportional to $q^2 \log{q^2}$ 
when the zero-momentum gluon propagator is constant ($\alpha_G=1$), as lattice data seems to points to 
(very recentely, the authors of \cite{Tissier:2010ts}, in a different context, have also found 
a ghost propagator dressing function whith the same low-momentum behaviour). 
The proportionality factor in front of it was also proven to be written in terms of the coupling at 
the renormalization momentum and the zero-momentum values of the gluon propagator and ghost dressing function. 
The aim of this note is to go further in the low-momentum analysis for the ghost 
propagator behaviour in that decoupling case with $\alpha_G=1$. With this purpose, 
a simple model for a massive gluon propagator (where the 
gluon mass is taken not to run with the momentum and to be approximated by its 
zero-momentum value) is applied in order to compute the ${\cal O}(q^2)$-correction for the 
low-momentum ghost dressing function. Then, we prove that this low-momentum behaviour 
is controlled by that gluon mass and by the zero-momentum value of the 
effective charge defined from the Taylor-scheme ghost-gluon vertex in 
ref.~\cite{Aguilar:2009nf} (see section \ref{revisiting}).
We also show this low-momentum formula to describe pretty well 
some lattice ghost dressing function data~\cite{Bogolubsky:2007ud,Boucaud:2009sd} for
different volumes and $\beta$'s. 
Some details of the computations are also provided in two appendices.

\Section{The ghost propagator Dyson-Schwinger equation}\label{revisiting}
We will start by following ref.~\cite{Boucaud:2008ky} and examine the Dyson-Schwinger equation for the ghost
propagator (GPDSE) which can be written diagrammatically as

\vspace{\baselineskip}
\begin{small}
\bea
\left(\ghostDr\right)^{-1}%
=
\left(\ghostBr\right)^{-1}%
- 
\ghostSD %
\nonumber
\eea\end{small}%

\noindent that, after omitting colour indices and dividing 
both sides by $k^2$, in Landau gauge reads
\begin{equation}
\label{SD1}
\begin{split}
\frac{1}{F(k^2)} & = 1 + g_0^2 N_c \int \frac{d^4 q}{(2\pi)^4} 
\left( \rule[0cm]{0cm}{0.8cm}
\frac{F(q^2)\Delta((q-k)^2)}{q^2 (q-k)^2} 
\left[ \rule[0cm]{0cm}{0.6cm}
\frac{(k\cdot q)^2}{k^2} - q^2  
        \right]
\ H_1(q,k)
           \right) \ ;
\end{split}
\end{equation} 
where $F$ stands for the ghost dressing function and 
$\Delta$ for the full gluon propagator form factor, 
\beq
\langle \widetilde{A}_\mu^a(k) \widetilde{A}_\nu^a(-k) \rangle = g^{T}_{\mu\nu}(k) \delta^{ab} \Delta(k^2) \ ,
\eeq
with $k^2 g^{T}_{\mu\nu}(q)= k^2 g_{\mu\nu}-k_\mu k_\nu$. The standard tensor decomposition for 
the ghost-gluon vertex, 

\beq
\widetilde{\Gamma}_\nu^{abc}(-q,k;q-k) \ &=& \ i g_0 f^{abc} q_{\nu'}  
\widetilde{\Gamma}_{\nu'\nu}(-q,k;q-k) \nonumber \\
&=&
i g_0 f^{abc} \left( \ q_\nu H_1(q,k) + (q-k)_\nu H_2(q,k) \ \right) \ ,
\label{DefH12}
\eeq
is applied, where $q$ and $k$ are respectively the outgoing and incoming ghost momenta and $g_0$ 
is the bare coupling constant. It should be noticed that, because of the transversality condition, $H_2$ defined in 
eq.~(\ref{DefH12}) does not contribute for the GPDSE in the Landau gauge. 
The integral equation (\ref{SD1}) is written in terms of bare Green functions. 
However, this equation is only meaningful after the specification of some appropriate UV-cutoff $\Lambda$, 
for instance: $F(k^2) \rightarrow F(k^2,\Lambda)$.
It can be cast into a renormalized form by dealing properly with UV divergencies, {\it i.e.}
\beq
g_R^2(\mu^2) &=& Z_g^{-2}(\mu^2,\Lambda) g_0^2(\Lambda) \nonumber \\
k^2 \Delta_R(k^2,\mu^2) &=& Z_3^{-1}(\mu^2,\Lambda) G(k^2,\Lambda) \nonumber \\
F_R(k^2,\mu^2) &=& \widetilde Z_3^{-1}(\mu^2,\Lambda) F(k^2,\Lambda) \ ,
\label{Ren}
\eeq
where $\mu^2$ is the renormalization momentum and $ Z_g, Z_3$ and $\widetilde Z_3$ the renormalization constants for the coupling, the gluon and the ghost respectively. $ Z_g$ is related to the ghost-gluon vertex renormalization constant (defined by
 $\widetilde{\Gamma}_R=\widetilde Z_1 \Gamma_B$) through $ Z_g= \widetilde{Z_1} (Z_3^{1/2}\,\widetilde Z_3)^{-1}$. Then  
Taylor's well-known non-renormalization theorem, which states that $H_1(q,0)+H_2(q,0)=1$ in Landau gauge and 
to any perturbative order, can be invoked to conclude that 
$\widetilde Z_1$ is finite. Thus, 
\beq\label{SDRnS}
\frac 1 {F_R(k^ 2,\mu^2)} \ = \ \widetilde Z_3(\mu^2,\Lambda) 
+ N_C \widetilde Z_1 \ g_R^2(\mu^2) \ \Sigma_R(k^2,\mu^2;\Lambda)  
\eeq
 where
\beq
\Sigma_R(k^2,\mu^2;\Lambda) &=& \int^{q^2 < \Lambda^2} \frac{d^4 q}{(2\pi)^4} 
\nonumber \\ 
&\times&
\left( \rule[0cm]{0cm}{0.8cm}
\frac{F_R(q^2,\mu^2)\Delta_R((q-k)^2,\mu^2)}{q^2 (q-k)^2} 
\left[ \rule[0cm]{0cm}{0.6cm}
\frac{(k\cdot q)^2}{k^2} - q^2  
        \right]
\ H_{1,R}(q,k;\mu^2) \right) \ . \nonumber \\
\label{sigma}
\eeq 
One should notice that the UV cut-off, $\Lambda$, is still required as an upper integration bound 
in eq.~(\ref{sigma}) since the integral is UV-divergent, behaving as 
$\int dq^2/q^2 (1+11 \alpha_S/(2\pi) \log{(q/\mu)}))^{-35/44}$. In fact, this 
induces a cut-off dependence  in $\Sigma_R$ that cancels against the one 
of $\widetilde Z_3$ in the r.h.s. of eq.~(\ref{SDRnS}), as can be easily seen by checking  
that $\widetilde Z_3^{-1}(\mu^2,\Lambda) \Sigma_R(k^2,\mu^2;\Lambda)$ approaches  some finite limit 
as $\Lambda \to \infty$ since the ghost and gluon propagator anomalous dimensions and the
beta function verify the relation $2 \widetilde \gamma + \gamma + \beta = 0$.
This is in accordance with the fact that the l.h.s. does not 
depend on $\Lambda$. Then, 
we will apply a MOM renormalization prescription: this means that all the Green functions 
take their tree-level value at the renormalization point,
\beq
F_R(\mu^2,\mu^2) \ = \ \mu^2 \Delta_R(\mu^2,\mu^2) \ = \ 1 \ .
\eeq
The subtraction point can be taken at any non-zero scale, $\mu^2$; however 
we prefer it to lie on the UV momentum domain (to have to deal, for instance, with renormalization 
constants or a renormalized coupling, $g_R(\mu^2)$, in \eq{SDRnS} that could be estimated from 
perturbation theory). We will also choose to 
renormalize the ghost-gluon vertex at the Taylor-theorem kinematics 
({\it i.e.}, a vanishing incoming ghost momentum), thus
\beq
\widetilde{Z}_1(\mu^2) \underbrace{\left( H_1(q,0) + H_2(q,0) \right)
}_{\displaystyle 1} \ = \ 1 \ .
\eeq
Now, in the following, $H_1(q,k)$ will 
be approximated by a constant with respect to both 
momenta and our MOM prescription implies thus
\beq
H_{1,R}(k,q;\mu^2) \ = \ \widetilde{Z}_1 H_1(k,q) \ = \ H_1 \ ,
\eeq
where $H_1$ is the assumed-to-be constant bare ghost-gluon vertex.

Although we cannot forget that the UV cut-off dependences in both sides 
of eq.~(\ref{SDRnS})  match only in virtue of the previously mentionned relation between 
the ghost and gluon propagator anomalous dimension and the beta function, 
in order  not to have to deal with the UV cut-off, we can procceed as follows:
we consider eq.~(\ref{SDRnS}) for two different scales, $k$ and $p$, such that
$p^2-k^2=\delta^2 k^2$ ($\delta$ being an extra parameter that, 
for the sake of simplicity, will be taken to be small enough as to 
expand on it around 0) and subtract them 
\beq
\frac{1}{F_R(k^2,\mu^2)} - \frac{1}{F_R(p^2,\mu^2)}  
\ = \  
N_C \ g_R^2(\mu^2) 
\left( \rule[0cm]{0cm}{0.5cm} \Sigma_R(k^2,\mu^2;\infty) - 
\Sigma_R(p^2,\mu^2;\infty) \right) \ . \nonumber \\
\label{SDRS}
\eeq
Then, the subtraction renders UV-safe the integral in the r.h.s. and the limit 
$\Lambda \to \infty$ can be explicitely taken,
\beq
\label{LamInf}
 \Sigma_R(k^2,\mu^2;\infty) - 
\Sigma_R(p^2,\mu^2;\infty) 
&=&  \ H_1 \ \int \frac{d^4 q}{(2\pi)^4} 
\left( \rule[0cm]{0cm}{0.8cm}
\frac{F(q^2,\mu^2)}{q^2} \left(\frac{(k\cdot q)^2}{k^2}-q^2\right) \right. 
\nonumber \\ 
 &\times& 
\left. 
\left[ \rule[0cm]{0cm}{0.6cm}
\frac{\Delta\left((q-k)^2,\mu^2\right)}{(q-k)^2} -  
\frac{\Delta\left((q-p)^2,\mu^2\right)}{(q-p)^2} 
\rule[0cm]{0cm}{0.6cm} \right]
\rule[0cm]{0cm}{0.8cm} \right) \ .
\eeq
An accurate analysis of eq.~(\ref{SDRS}) requires~\cite{Boucaud:2005ce}, in addition,  to cut the 
integration domain of eq.~(\ref{LamInf}) into two pieces by introducing 
some new scale $q_0^2$ ($q_0$ is a momentum scale below which the IR behaviour is a 
good approximation for both ghost and gluon),
\beq\label{q0}
\Sigma_R(k^2,\mu^2;\infty) - \Sigma_R(p^2,\mu^2;\infty) 
\ = \ 
H_1 \left( \rule[0cm]{0cm}{0.4cm} I_{\rm IR}(k^2) \ + \ I_{\rm UV}(k^2) \right)
\eeq
where $I_{\rm IR}$ represents the integral in eq.~(\ref{LamInf}) over $q^2 < q_{0}^2$ and 
$I_{\rm UV}$ over $q^2 > q_{0}^2$. We only wrote explicitly the dependence on $k^2$ for the r.h.s. 
because we shall expand on $\delta$ around zero with $\mu^2$ kept fixed. Then, 
for $k^2,p^2 \ll q_{0}^2$, we will propose the following ansatz\footnote{This is the massive gluon 
propagator where the gluon running mass~\cite{Lavelle:1991}, $M(q^2)$, appears to be approximated by 
its frozen value at vanishing momentum, $M(0)$.}: 
\beq\label{gluonprop}
\Delta_{\rm IR}(q^2,\mu^2) &\simeq& \frac{B(\mu^2)}{q^2 + M^2} 
\ = \ \frac{B(\mu^2)}{M^2} \left( 1 \ - \ \frac{q^2}{M^2} + \mathcal{O}\left(\frac{q^4}{M^4}\right) \right) \ ,
\eeq
for a massive gluon propagator that implies of course $\alpha_G=1$, as the 
current lattice data seems to point to. 
At this stage, we should remember that \eq{gluonprop}, or any other additional hypothesis about the 
low-momentum gluon behaviour, is needed to specify the ${\cal O}(q^2)$-correction (next-to-leading) for the ghost 
dressing function.
It should be also noted that, provided that the gluon propagator 
is to be multiplicatively renormalized, the mass scale, $M$, in \eq{gluonprop} 
does not depend on renormalization scale, $\mu^2$.
Now, we shall look for the ghost dressing function, $F_{\rm IR}$,
its leading behaviour being parameterized through a general power law behaviour, 
\beq\label{dress}
F_{\rm IR}(q^2,\mu^2) &=& A(\mu^2) \left( \frac{q^2}{M^2} \right)^{\alpha_F} \left( 1 + \cdots 
\rule[0cm]{0cm}{0.6cm} \right) \ ,
\eeq
where $\alpha_F > -2$ to keep the integral $I_{\rm IR}$ infrared convergent.
After some algebra (see appendix \ref{app:IR}), we obtain:
\beq\label{IRKi}
I_{\rm IR}(k^2) 
&\simeq& 
- \frac{\delta^2}{M^{2+2\alpha_F}} \frac{2 A(\mu^2) B(\mu^2)}{(2\pi)^3} 
 \sum_{i=0}^\infty \ (4 k^2)^i  C_i \ \displaystyle \int_0^{q_{0}} q^{3+2i+2\alpha_F} dq \
K_i(q^2;k^2,M^2) \ + \ {\cal O}(\delta^4) \nonumber \\
\eeq
where
\beq\label{Rser}
K_i(q^2;k^2,M^2) &=&   \frac {i} {(q^2+ k^2 + M^2)^{2i+1}} - \frac {i} 
{(q^2+ k^2)^{2i+1}}  \nonumber \\ 
&&  \ + \ \ k^2 \left( \frac {2i+1} {(q^2+ k^2)^{2i+2}} 
- \frac {2i+1} {(q^2+ k^2 + M^2)^{2i+2}} \right) 
\eeq
and
\beq
C_i \ = \ \frac{12 \pi^2 4^{i}}{\Gamma(-3/2-i) \Gamma(1/2-i) \Gamma(5+2i)} \ .
\eeq
From now on, we will focus on the decoupling case: $\alpha_F=0$. Then, 
as shown in appendix \ref{app:aF0}, the integral in \eq{IRKi} can be 
written as a series in powers of $k^2$, the leading term 
given by 
\beq
I_{\rm IR}(k^2) 
\ \simeq \ 
\delta^2 \ \frac{A(\mu^2) B(\mu^2)}{64\pi^2} \ 
\frac{k^2}{M^2} \ 
\left[ \ln{\frac{k^2}{M^2}} - \frac 5 6 
\ + \ {\cal O}\left(\frac{M^2}{q_0^2}\right) 
\right] \ + \ 
{\cal O}\left(\frac{k^4}{M^4},\delta^4\right)
\nonumber \\
\label{demoF}
\eeq
Then, the first correction to the leading constant term for the ghost dressing 
function should be
\beq\label{solFIR1}
F_{\rm IR}(q^2,\mu^2) \ = \ A(\mu^2) \left( 1 + 
A_2(\mu^2) \frac{q^2}{M^2} \left[ \ln{\frac{q^2}{M^2}} -  \frac {11} 6  \
\right] \
\ + \ \cdots \right) \ ,
\eeq 
such that
\beq
\frac 1 {F(k^2,\mu^2)} - \frac 1 {F(p^2,\mu^2)} \simeq
\delta^2 \frac{A_2(\mu^2)}{A(\mu^2)} \ \frac{k^2}{M^2} \ 
\left(
\ln{\frac{k^2}{M^2}} - \frac{5}{6} 
\right) 
+ {\cal O}\left(\frac{k^4}{M^4},\delta^4 \right) \ ,
\eeq
the \eq{SDRS} being satisfied when:
\beq
A_2(\mu^2) \ = \ N_C g^2_R(\mu^2) H_1 
\ \frac{A^2(\mu^2) B(\mu^2)}{64 \pi^2}
\eeq
Thus, up to corrections of the order of $k^4/M^4$, 
one shall have:
\beq\label{solFIR2}
F_{\rm IR}(q^2,\mu^2) = F_{\rm IR}(0,\mu^2) \left( 1 \ + \ \frac{N_C H_1 R}{16 \pi} \ 
q^2 \left[ \ln{\frac{q^2}{M^2}} - \frac{11}{6} \right]
\ + \ {\cal O}\left(\frac{q^4}{M^4} \right) \right)
\eeq
where:
\beq\label{coefC} 
R & = & \frac{g^2_R(\mu^2)}{4 \pi} 
F_{\rm IR}^2(0,\mu^2) \Delta_{\rm IR}(0,\mu^2) 
\ = \ \lim_{q^2->0} \frac{\alpha_T(q^2)}{q^2} 
\eeq 
It worth pointing that, provided that $g_R$ is renormalized in the 
Taylor scheme (the incoming ghost momentum vanishing in the renormalization 
point) $R$ is a $\mu$-independent (RGI) quantity\footnote{This claim is equivalent 
to that of ref.~\cite{Boucaud:2009sd} about the cut-off independence of the bare  
ghost-dressing-function subleading term.}, as it is manifest from 
\eq{coefC}, where $\alpha_T=g_T^2/(4\pi)$ is the perturbative strong coupling 
defined in this Taylor scheme~\cite{Boucaud:2008gn}. 
However, for phenomenological purposes a coupling vanishing at zero-momentum 
is not convenient and, instead of that, a non-perturbative 
effective charge is defined from the gluon propagator in ref.~\cite{Aguilar:2008fh}, 
within the framework of the pinching technique~\cite{Cornwall}, which can be appropriatedly 
extended to the Taylor ghost-gluon coupling~\cite{Aguilar:2009nf}.
As a consequence of the appropriate {\it amputation} of a 
massive gluon propagator,  where the gluon mass scale is the same RI-invariant mass scale appearing in 
\eq{gluonprop}, this Taylor effective charge is frozen at low-momentum and gives 
a non-vanishing zero-momentum value~\cite{Aguilar:2009nf}, 
\beq\label{coefC2}
\overline{\alpha}_T(0) = \lim_{q \to 0} \left(q^2 + M^2 \right) \frac{\alpha_T(q^2)}{q^2} 
\ = M^2 R \ ,
\eeq
in terms of which the ghost-dressing-function subleading correction can be expressed:
\beq\label{solFIRJo}
F_{\rm IR}(q^2,\mu^2) = F_{\rm IR}(0,\mu^2) \left( 1 \ + \ 
\frac{N_C H_1}{16 \pi} \ \overline{\alpha}_T(0) \ 
\frac{q^2}{M^2} \left[ \ln{\frac{q^2}{M^2}} - \frac {11} 6 \right]
\ + \ {\cal O}\left(\frac{q^4}{M^4} \right) \right)
\eeq

\noindent
It should be also noted that eqs.~(\ref{solFIR1},\ref{solFIR2}) imply to 
take $M^2/q_0^2 \ll 1$, as it is manifest from \eq{demoF}. However, 
any correction to that approximation will not play at the order of 
the coefficient eqs.~(\ref{coefC},\ref{coefC2}), that will keep the same 
value disregarding that of $M^2/q_0^2$, but at the order of the gluon 
mass, $M^2$, inside the logarithm (exactly like the factor 5/6 in \eq{demoF}).

\Section{Comparison with ghost propagator lattice data}

In the last few years, many works have been devoted, at least partially, to the computation of 
the ghost propagator by using lattice simulations. In ref.~\cite{Boucaud:2009sd}, some of those ghost propagators 
results were collected, mainly the ones for big lattice volumes from ref.~\cite{Bogolubsky:2007ud}, and 
studied in a different context but shown to verify the asymptotic low-momentum expansion for the 
ghost propagator where only the leading term, $q^2 \log(q^2)$, was 
kept (see fig.~3 of ref.~\cite{Boucaud:2009sd}). 

\begin{figure}[hbt]
\begin{center}
\includegraphics[width=12cm]{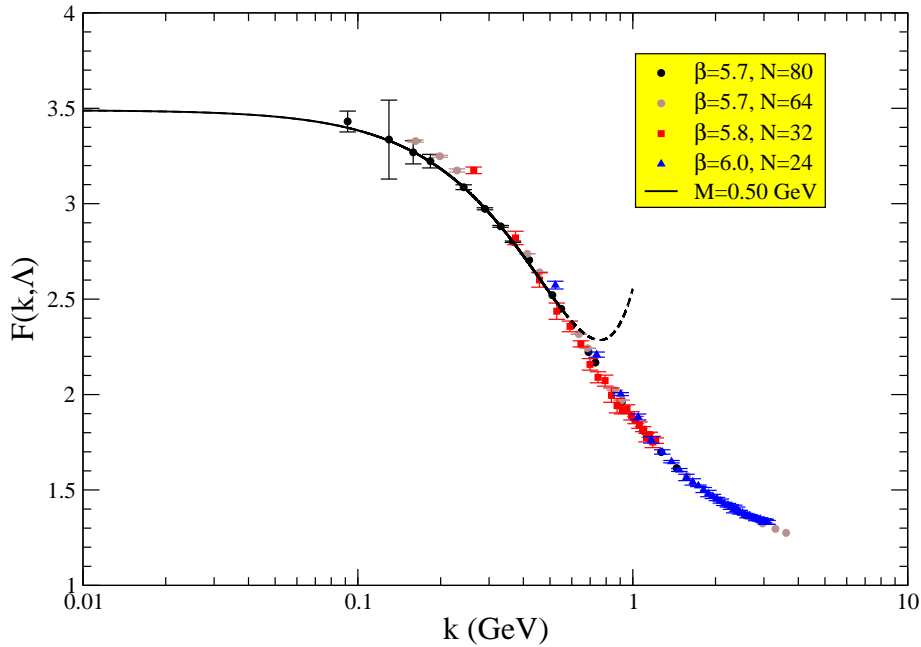} 
\end{center}
\caption{\small Ghost dressing function from lattice data~\cite{Bogolubsky:2007ud,Boucaud:2009sd} pretty well described by 
the low-momentum formula, \eq{solFIRJo}, with a $R=10(1)$ GeV$^{-2}$ from ~\cite{Boucaud:2009sd} 
and a best-fit for $M=0.50(2)$ GeV.}
\label{fig:all-gh}
\end{figure}

Now, we will consider among the results collected in ref.~\cite{Boucaud:2009sd} those for the bigger 
lattice volumes and confront them to \eq{solFIR2} or \eq{solFIRJo}, where the ${\cal O}(q^2)$-corrections have 
been incorporated. Thus, the low-momentum behaviour of the ghost dressing function being 
determined by the gluon mass, $M$, and the zero-momentum effective charge, 
$\overline{\alpha}_T(0)$, they could be obtained by fitting \eq{solFIR2} to the lattice data 
and applying \eq{coefC}. However, as previously pointed, $R=\overline{\alpha}_T(0)/M^2$ is 
a RGE-quantity that can be directly obtained from bare lattice ghost and gluon propagators. 
The latter is done in ref.~\cite{Boucaud:2009sd}, in particular at $\beta=5.7$ for a $80^4$ lattice
(precisely exploiting the lattice data from ref.~\cite{Bogolubsky:2007ud}) and $R \simeq 10.$ GeV$^{-2}$ is obtained.
Thus, we will fit \eq{solFIRJo}, where we approximate $H_1=1$ and take $\overline{\alpha}_T(0)/M^2 = 10$ GeV$^{-2}$, 
to the ghost propagator lattice data and obtain the 
curve plotted in fig.~\ref{fig:all-gh} for the best-fit parameters given in tab.~\ref{tab:all-gh}.

\begin{table}[hbt]
\begin{center}
\begin{tabular}{||c||c|c|c||}
\hline
 & $R=\overline{\alpha}_T(0)/M^2$ (GeV$^{-2}$)& $M$ (GeV) & $\overline{\alpha}_T(0)$ \\
\hline
$\beta=5.7(80^4)$ & 10(1) & 0.50(2) & 2.5(3) \\
\hline
\end{tabular}
\end{center}
\caption{\small Best-fit parameters obtained by describing the ghost dressing function lattice data 
with \eq{solFIRJo} (see fig.~\ref{fig:all-gh}). The errors quoted do not account for any systematical 
uncertainty.}
\label{tab:all-gh}
\end{table}

Thus, the ghost propagator lattice data behave pretty well as \eq{solFIRJo} asks for with a 
gluon mass, $M=0.50(2)$ GeV, in the right ballpark (roughly from 400 MeV to 700 MeV) 
defined by phenomenological tests~\cite{Halzen:1992vd} or direct lattice measurements from 
the gluon propagator~\cite{Bonnet:2001uh}. 
It should be emphasized that, the RGI quantity $R$ being 
determined by vanishing-momentum ghost and gluon propagators~\cite{Boucaud:2009sd}, the 
only parameter controlling the functional behaviour of the ghost propagator to 
be fitted is the gluon mass. 
However, it should be remembered that ${\cal O}(M^2/q_0^2)$-corrections in \eq{demoF} 
will play at the order of ${\cal O}(q^2)$ in \eq{solFIRJo} and, although not modifying 
the low-momentum functional behaviour, the fitted gluon mass can be borrowing 
something from these corrections. Consequently, the latter prevents 
us to take that fitted gluon mass as a precise determination but as an approximative value 
that indeed appears to be in the very right ballpark.

In summary, the low-momentum ghost propagator dressing 
function computed from big-volume lattices can be very well described by the asymptotical 
formula \eq{solFIRJo} with a value for the vanishing-momentum effective charge, 
$\overline{\alpha}_T(0)\simeq 2.5$, pretty well in agreement with its direct lattice determinations 
as the gluon mass, $M\simeq 0.5$ GeV, lies on its phenomenological range.

\medskip

\Section{Conclusions}\label{conclu}

The ghost propagator DSE, with the only assumption of taking $H_1(q,k)$ from the 
ghost-gluon vertex in \eq{DefH12} to be constant in the infrared domain of $q$, can be 
exploited to look into the low-momentum behaviour of the ghost propagator.  The
two classes of solutions named ``decoupling'' and ``scaling'' can be indentified and 
shown to depend on whether the ghost dressing function achieves a finite non-zero
constant ($\alpha_F=0$) at vanishing momentum or not ($\alpha_F \neq 0$). 
In accordance with the fact that the lattice simulations indicates that the gluon propagator is 
finite and non-vanishing at zero momentum, we applied in this paper a model with 
a massive gluon for the infrared gluon propagator to obtain the low-momentum 
behaviour of the ghost propagator. We focussed on $\alpha_F=0$ (decoupling) and derive 
an asymptotic expression reliable up to ${\cal O}(q^2)$ for the low-momentum 
ghost propagator. This low-momentum behaviour results to be regulated by the 
gluon propagator mass and by a regularization-independent 
dimensionless quantity that appears to be the effective charge defined from 
the Taylor-scheme ghost-gluon vertex at zero momentum.
Finally, this asymptotic expression is also proven to fit 
pretty well the low-momentum ghost propagator data obtained from very big lattices 
simulations with a gluon mass, $M \sim 500$ MeV, that appears to lie on the 
right ballpark.

\paragraph{Acknowledgements: } JRQ and MEG acknowledge support from the research projects P07FQM02962 funded 
by ``Junta de Andaluc\'ia'', FPA2009-10773 and Consolider-Ingenio   
CSD2007-00042 funded by the Spanish MICINN.

\appendix 

\Section{ The integral $I_{\rm IR}$} 

The integral $I_{\rm IR}$ is defined as:
\beq
\label{app:IR}
I_{\rm IR}(k^2)  
&=&  \int_{q^2<q_0^2} \frac{d^4 q}{(2\pi)^4} 
\left( \rule[0cm]{0cm}{0.8cm}
\frac{F_{\rm IR}(q^2,\mu^2)}{q^2} \left(\frac{(k\cdot q)^2}{k^2}-q^2\right) \right. 
\nonumber \\ 
 &\times& 
\left. 
\left[ \rule[0cm]{0cm}{0.6cm}
\frac{\Delta_{\rm IR}\left((q-k)^2,\mu^2\right)}{(q-k)^2} -  
\frac{\Delta_{\rm IR}\left((q-p)^2,\mu^2\right)}{(q-p)^2} 
\rule[0cm]{0cm}{0.6cm} \right]
\rule[0cm]{0cm}{0.8cm} \right) \ .
\eeq
where, as shown in eqs.~(\ref{gluonprop},\ref{dress}), we take:
\beq\label{app:gluonpar}
\Delta_{\rm IR}(q^2,\mu^2) &\simeq& \frac{B(\mu^2)}{q^2 + M^2} \ ,
\\ \label{app:ghostpar}
F_{\rm IR}(q^2,\mu^2) &\simeq & A(\mu^2) \left( \frac{q^2}{M^2} \right)^{\alpha_F} \ .
\eeq
Provided that $\alpha_F > -2$, the integral $I_{\rm IR}$ shall be infrared convergent and 
one then obtains:
\bea
I_{\rm IR}(k^2) 
&\simeq& 
-  \frac{2 A(\mu^2) B(\mu^2)}{(2\pi)^3 M^{2\alpha_F}} 
\displaystyle \int_0^{q_{0}} q^{3+2\alpha_F} dq \ \int_0^\pi \sin^4{\theta} \ d\theta \          
\nonumber \\ 
&\times& 
\left( \frac 1 {q^2 + k^2 - 2kq \ \cos\theta} 
\ \frac 1 {q^2+ k^2 + M^2 -2kq \ \cos\theta} 
\right.
\nonumber \\
& & - \left.
\frac 1 {q^2 + p^2 - 2kq \ \cos\theta} 
\ \frac 1 {q^2+ p^2 + M^2 -2kq \ \cos\theta}
\right)
\label{I1}
\\
&\simeq& 
-  \rule[0cm]{0cm}{0.95cm} \frac{1}{M^{2(1+\alpha_F)}} \frac{2 A(\mu^2) B(\mu^2)}{(2\pi)^3} 
\displaystyle \int_0^{q_{0}} q^{3+2\alpha_F} dq \ \int_0^\pi \sin^4{\theta} \ d\theta \          
\nonumber \\ 
&\times& 
\left( 
\frac{1}{q^2 + k^2 - 2kq \cos\theta}
- \frac{1}{q^2 + p^2 - 2pq \cos\theta}
\right.
\nonumber \\
& &  \left. - \ 
\frac{1}{q^2 + k^2 + M^2- 2kq \cos\theta}
+ \frac{1}{q^2 + p^2 + M^2 - 2pq \cos\theta}
\right) \ .
\eea
We now apply that:
\beq
\int_0^\pi \sin^4{\theta} \ d\theta \ 
\frac 1 {q^2 + k^2 - 2kq \ \cos\theta} 
\ = \
\frac 1 {q^2 + k^2} \
\sum_{i=0}^\infty 
\left( \frac{2 k q}{q^2 + k^2} \right)^{2i} \
\underbrace{ \int_0^\pi d\theta \ \sin^4{\theta} \cos^{2i}\theta
}_{\displaystyle C_i} \ ,
\eeq
where we have taken into account that angular integral vanishes for 
odd powers of the $\cos$. The coefficients $C_i$ can be analytically 
obtained:
\beq
C_i \ = \int_0^\pi d\theta \ \sin^4{\theta} \cos^{2i}\theta \ = \ 
\frac{12 \pi^2 4^{i}}{\Gamma(-3/2-i) \Gamma(1/2-i) \Gamma(5+2i)} \ .
\eeq
Then, one can write:
\beq
I_{\rm IR}(k^2) 
&\simeq& 
-  \frac{1}{M^{2+2\alpha_F}} \frac{2 A(\mu^2) B(\mu^2)}{(2\pi)^3} 
 \sum_{i=0}^\infty \ (4 k^2)^i  C_i \ \displaystyle \int_0^{q_{0}} q^{3+2i+2\alpha_F} dq \
\\
&\times&
\underbrace{\left( 
\frac{1}{(q^2+k^2)^{2i+1}} \ - \ 
\frac{(1+\delta^2)^{i}}{(q^2+p^2)^{2i+1}} \ + \
\frac{(1+\delta^2)^{i}}{(q^2+p^2+M^2)^{2i+1}} \ - \
\frac{1}{(q^2+k^2+M^2)^{2i+1}}
\right)}_{\displaystyle R}
\nonumber
\eeq
and thus expand in terms of $\delta$,
\beq\label{ap:Rser}
R &=& \delta^2 \ \left[ \frac {i} {(q^2+ k^2 + M^2)^{2i+1}} - \frac {i} 
{(q^2+ k^2)^{2i+1}} \right. \nonumber \\ 
&& \left. \ + \ \ k^2 \left( \frac {2i+1} {(q^2+ k^2)^{2i+2}} 
- \frac {2i+1} {(q^2+ k^2 + M^2)^{2i+2}} \right) \right] \ + \ {\cal O}(\delta^4) \ ,
\eeq
to obtain \eq{IRKi}.

\bigskip

\Section{The case $\alpha_F=0$}
\label{app:aF0}

When $\alpha_F=0$, the integral $I_{\rm IR}$ in \eq{IRKi} can be expanded as a series 
on powers of $k^2$ leaded by the following term:
\beq
I_{\rm IR}(k^2) 
&\simeq& 
- \delta^2 \ \frac{2 A(\mu^2) B(\mu^2)}{(2\pi)^3} \ \frac{1}{M^2} \
\times
\\
&& \sum_{i=0}^\infty (4 k^2)^i \ C_i \ \int_0^{q_{0}} q^{3+2i} dq \ 
\left(
\frac {i} {(q^2+ k^2 + M^2)^{2i+1}} - 
\frac {i} {(q^2+ k^2)^{2i+1}} \right.
\nonumber \\
&& \left. +  \frac {(2i+1) k^2} {(q^2+ k^2)^{2i+2}} 
- \frac {(2i+1)k^2} {(q^2+ k^2 + M^2)^{2i+2}} \right)
\ + \ {\cal O}(\delta^4)
\nonumber \\
\rule[0cm]{0cm}{1cm} &\simeq& 
- \delta^2 \ \frac{A(\mu^2) B(\mu^2)}{(2\pi)^3} \ \frac{k^2}{M^2} \
\times
\nonumber \\
&& \sum_{i=0}^\infty 4^i \ C_i
\int_0^{\infty} dt \ t^{1+i}
\left( 
\frac {i} {(1+ t + \frac{M^2}{k^2})^{2i+1}} - 
\frac {i} {(1+ t)^{2i+1}} \right.
\nonumber \\
&& \left. +  \frac {2i+1} {(1+ t)^{2i+2}} 
- \frac {2i+1} {(1 + t + \frac{M^2}{k^2})^{2i+2}} \right)
\ + \ \cdots
\label{ap:demoF}
\eeq
Then, by integrating in \eq{ap:demoF} and 
expanding consistently in terms of $k^2/M^2$, one obtains:
\beq
I_{\rm IR}(k^2) 
&\simeq& \rule[0cm]{0cm}{1.1cm} 
\delta^2 \ \frac{A(\mu^2) B(\mu^2)}{64\pi^2} \ 
\frac{k^2}{M^2} \ 
\left[ \ln{\frac{k^2}{M^2}} - \frac 5 6 
\ + \ {\cal O}\left(\frac{M^2}{q_0^2}\right) 
\right] \ + \ 
{\cal O}\left(\frac{k^4}{M^4},\delta^4\right) \ ,
\eeq
which is the \eq{demoF} that gives the result for $I_{\rm IR}$ in the 
regular case.



%
%
\end{document}